\documentclass[final]{cimento}
\input psfig.sty
\begin{document}

%\begin{titlepage}
%$\ $

\title{
On the Extensive Air Shower density spectrum}

\author {
Aleksander Zawadzki \from{cdf},
Tadeusz Wibig \from{ul},
\atque
Jerzy Gawin \from{ipj}}
\instlist {
\inst{cdf}
Coll\`ege de France, Paris
\inst{ul}
Experimental Physics Dept., University of
\L \'{o}d\'{z},
Pomorska 149/153, 90-236
\L \'{o}d\'{z},
Poland
\inst {ipj}
Soltan Institute of Nuclear Studies,
Uniwersytecka 5, 90-950
\L \'{o}d\'{z},
Poland
}
\PACSes{\PACSit
{96.40.Pq}{Extensive air showers.}}

%\date{\today}

%\maketitle
%\thispagestyle{empty}
\maketitle
%\newpage
\begin{abstract}
In search for new methods of determining the primary energy spectrum
of Cosmic Rays, the attention was paid to the density spectrum measurement.
New methods available at present warrant an accurateness of conclusions
derived from the density spectrum measurements.
The general statement about the change of the spectral index of the
charged particle density spectrum is confirmed very clearly.
Results concerning the shower size and primary energy spectra are
also presented and discussed.
Interesting future prospects for applications of the density spectrum method
are proposed.
\end{abstract}
\section{Introduction}

The small array of Geiger-M\"uller (G-M) counters triggered by some coincidence of them
gives a very well defined information about the charged particle density at a
certain point in the Extensive Air Shower (EAS) plane. The binary type
information (nothing registered - at least one particle traversed the detector
sensitive area) lead to the limitations of the physical information but, on
the other hand, the measurement is free from additional uncertainties
introduced always by
the amplitude sensitive counters. What is also important the small grammage of
G-M counters assure that the charged particles from EAS are what is really
measured. This is specially important for a proper taking into account the
trigger effects and all the biases close to the small density edge of the
measurement domain.

Forty years ago the density spectrum measurements were in the focus of
interest
of Cosmic Ray (CR) physics. The power--law shape of the spectrum and the
behaviour of its spectral index were widely discussed (see e.g.
Ref.\ \cite{general}). The exact
measurement with high statistics and a careful statistical analysis was
performed and the results were given in Ref.\ \cite{az} by A. Zawadzki.
There was
shown clearly the steepening of the density spectrum in the relatively wide
range of densities around $\sim ~10~{\rm m}^{-2}$.

In the present paper we would like to re--examine the same experimental data
with the help of much more powerful methods available now. Some assumptions
necessary to obtain results in an analytical way are confirmed by numerical
calculations.

In the second part of this work we would like to proceed a bit further. The
general purpose is to use the old data set to get a new information about the
shower size spectrum or even primary particle energy spectrum.

The original work (Ref.\ \cite{az}) was limited to the charged particle density
spectrum problem only. It was obtained as a pure experimental result without
any assumptions concerning the EAS picture. The step we want to follow at
present is of much more delicate nature because anyway it is relied on the
charged particle spatial distribution in the shower plane or even on the
knowledge of the primary particle mass and the high energy interaction model,
respectively for the size spectrum and for the primary particle energy spectrum
case. The only possible way of solving the problem is to use the Monte--Carlo
shower simulation calculations believing in their correctness (to some extent,
as
it will be discussed). For the present work the well--known code CORSIKA v4.12
Ref.\ \cite{cors} has been used.

Finally we want to emphasize that the newly obtained density spectrum confirms
the calculations first published forty years ago.
%The size and energy spectra
%recovered from the same data are consistent with the general knowledge in a
%certain region of the CR domain.

Some short discussion of interesting feature seen for very small particle
densities will be given at the end and some future possibilities of using the
density spectrum method will be briefly announced.

\section{Data}
The data were collected in 1954--56 years using the three G--M counter
stations each of 18 tubes of 3 cm \o \  and 0.013 ${\rm m}^2$ sensitive
surface. The stations were placed in vertices of a rectangular triangle with
catheti of about 8 m. Two of them were used only for triggering (at
least one of G--M counters registered something) and the number of G--M signals
from the third station was used for further analysis. The question of the
efficiency and stability of the whole apparatus was widely discussed in
Ref.\ \cite{az} where it was shown its extremely exact behaviour during all the
time the measurement has been performed. The number of registrations (showers
registered) was 227712. For the comparison: three previous experiments were
made with 3039 (Ref.\ \cite{broa}), 8177 (Ref.\ \cite{hods}) and 22450 (Ref.\
\cite{coco}) statistics and the one of most accurate measurement
made at Moscow State University (Ref.\ \cite{ch})
more than twenty years later consists of 11500 events
(but for higher density range).

\begin{figure}
\centerline{\psfig{file=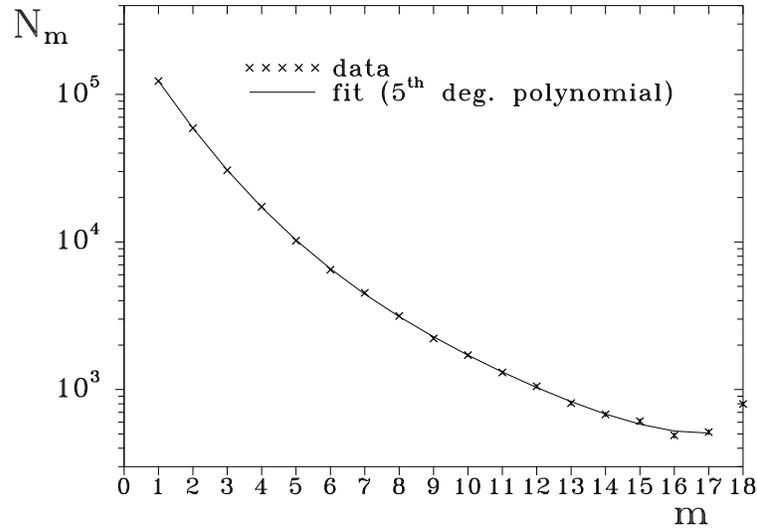,width=10cm}}
\caption{ Number of events $N_{\rm m}$ registered in the experiment with
a given value of m G--M tubes fired. The line is the $5^{\rm th}$ order
polynomial fit to the data in the ($\log N_{\rm m}$ -- m) scale.)}
\end{figure}

The data are presented in the Fig. 1. For the small values of m (number of the
G--M counters which registered a particle passing through) the statistical
accuracy is extremely high. Even for the large showers (big m values) errors
are of order of percents. The $5^{\rm th}$ order polynomial fit in the
$\log N_{\rm m}$
($N_{\rm m}$ is the number of events with a particular value of
m G--M tubes fired)
describe the data ( for $ 1 \le {\rm m} \le 17$) very well (with $\chi ^2$/NDF
of order of 1). As it was mentioned in Ref.\ \cite{az} to study the
dependence of
$N_{\rm m}$ with respect0to m the uncertainty at a particular value of m does
not
correspond to the statistical measurement error of the point at m alone. In
that sense all the points are not uncorrelated. The proper treatment of the
accuracy was discussed in Ref.\ \cite{az} and will be briefly presented
later on.

\begin{figure}
\centerline{\psfig{file=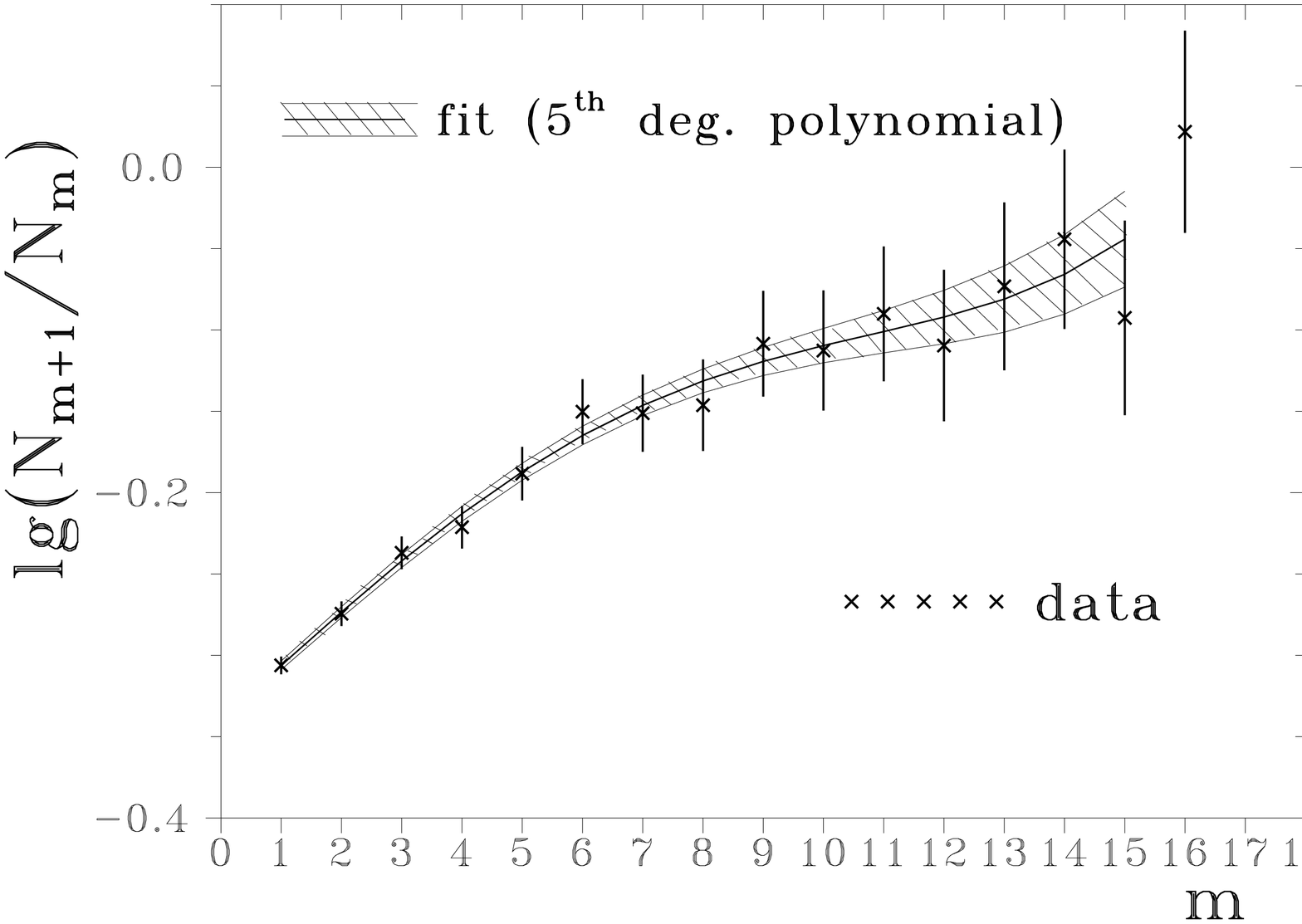,width=10cm}}
\caption{The ratio of $N_{\rm m+1}/N_{\rm m}$ as a function of m. The line
shows the results of the fit in the Fig. 1. The dashed area represents the
1$\sigma$ significance fit error.}
\end{figure}

Due to the problems connected with the interpretation of these data the more
convenient presentation is given in the Fig. 2. The ratio of the consequent
m-hit rates is sensitive only to the mean value of the first (and higher orders
of course) derivative of the charged particle density spectrum in a relatively
small density range, while the m-hit rate ($N_{\rm m}$) depends on the shape of the
spectrum as a whole. This point will be discussed below. The line in the Fig. 2
represents the polynomial fit from the Fig. 1. The errors given separately
for each point are statistical only. The dashed area is the
1$\sigma$ significance level of the fit so it represents the true error of the
measurement.

\section{The analysis}

\subsection{The density spectrum}
As it was briefly discussed the G--M hodoscopic measurement gives the
statistically clear basis to evaluate the true charged particle density
spectrum. Because the trigger detectors were rather close (in the comparison
with the characteristic dimension of the EAS particle lateral distribution) it
is quite reasonable to assume that in every case the EAS particle densities in
all three G--M stations were exactly the same. On the other hand the spacing
between stations was big enough to assume that the particular numbers of
particles passing each array fluctuate independently. These two assumptions
allow us to calculate the probability that a given particle density leads to
the registration of the event:

\begin{equation}
q(\rho) {\rm d}\rho~=~ {\rm d}\rho~
\left(1 - {\rm e}^{-sM\rho}\right)^2~f(\rho)~~~,
\end{equation}

\noindent
where: $s~=~0.013 {\rm m}^2$ is a single G--M tube area, $M~=~18$ is a number
of G--M tubes in each station and $f(\rho)$ is the density spectrum - the point
of our interest.

If we denote by $p({\rm m},\rho)$ the probability that for a given particle
density m of G--M counters register at least one particle then:

\begin{equation}
P_{\rm m}~=~ \int_{0}^{\infty} {\rm d}\rho~q(\rho)~p({\rm m},\rho)~~~.
\end{equation}

The lack of any correlation between EAS particles is assumed.
This possibility is a common one but not very obvious
(it will be discussed later).
Then
$p({\rm m},\rho)$ is given as a binomial distribution (with m = 0 excluded due
to the triggering condition).

\begin{figure}
\centerline{\psfig{file=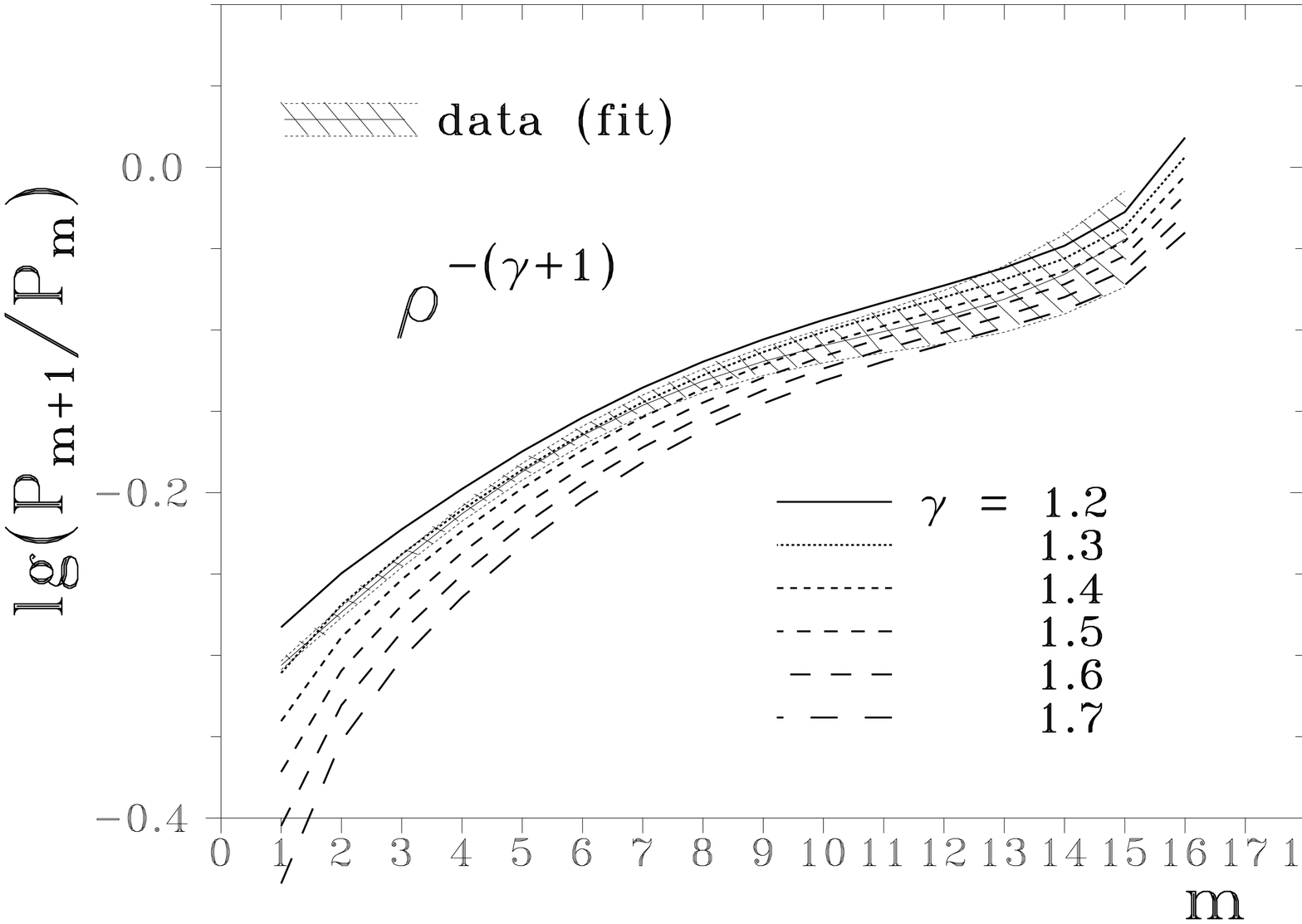,width=10cm}}
\caption{The ratio of $P_{\rm m+1}/P_{\rm m}$ as a function of m calculated
using Eq.\ (3) for different constant values of $\gamma$. The line shows the
experimental result as it was given in the Fig. 1.}
\end{figure}

The first approximation of the density spectrum: the power--law function is
used. The aim of the work is to test if the index of this spectrum - $\gamma$
is constant and to determine its eventual change as the particle density
changes. So we finally get normalized to the unity m-hit counting
rate:

\begin{equation}
P_{\rm m}~=~ \int_{0}^{\infty} {\rm d}\rho~\left(1 - {\rm e}
^{-sM\rho}\right)^2~
{ m \choose M } ~\left(1 - {\rm e}^{-s\rho}\right)^{\rm m}~
{\rm e}^{-s(M-{\rm m})\rho}~C~\rho ^{-(\gamma(\rho)+1)}
~~~,
\end{equation}

\noindent
where $C$ is a spectrum normalization constant.

The Eq.\ (3) can be evaluated analytically (for constant $\gamma$) for
each integer value of m as it was performed in Ref.\ \cite{az} in a rather
sophisticated manner. The numerical integration confirms the exactness of
forty years old calculations. Results show clearly that it is not possible to
describe the data with the one constant value of spectrum index. The
presentation of $P_{\rm m}$ for different $\gamma$'s however contains a
normalization constant which can be {\it a priori} different for each value of
$\gamma$ tested. To avoid the confusion in the Fig. 3 the values of
$P_{\rm m+1}/P_{\rm m}$ are presented.

For such ratios the problem of
normalization does not exist (at least if one assumes the change in $\gamma$
to be not a very dramatic one as is confirmed by the calculations). Detailed
analysis of this figure allows us to describe the change of the density
spectrum index as a function of m - the number of G--M tubes with at least one
charged particle seen in the event. To obtain the change of the $\gamma$
with the particle density ($\gamma(\rho)$ as it is in Eq.\ (3)) the
dependence of
the mean density contributing to a particular m can be calculated as:

\begin{equation}
f_{\rm m}(\rho) ~=~ q(\rho)~p({\rm m},\rho)~\sim~p({\rm m},\rho)
~\rho ^{-(\gamma(\rho)+1)}~{\left(1 - {\rm e}^{-sM\rho}\right)}^2~~~.
\end{equation}

The results are given in the Fig. 4

\begin{figure}
\centerline{\psfig{file=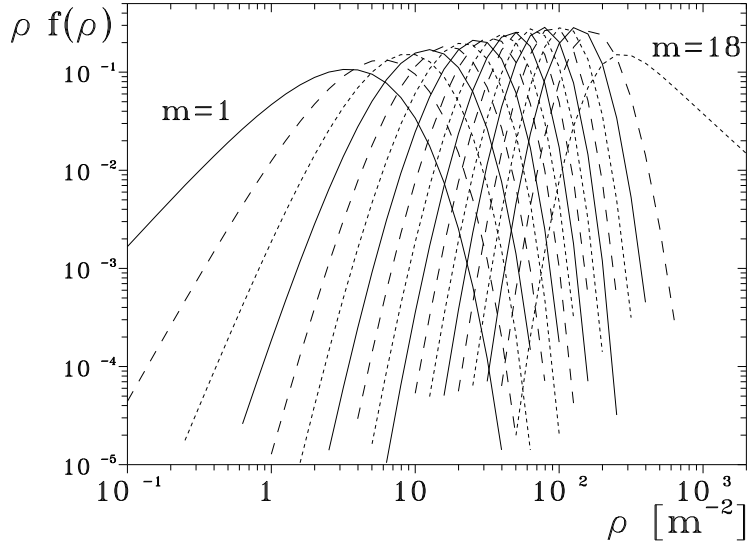,width=10cm}}
\caption{The distribution of particle
densities giving a contribution to the m-hit rate.}
\end{figure}

As it was expected the range of densities contributing to particular
$f_{\rm m}$ (except the case m = 18) is rather narrow so we can use the mean
value of each $f_{\rm m}(\rho)$ as an average density respected somehow to the
value of m.

\begin{figure}
\centerline{\psfig{file=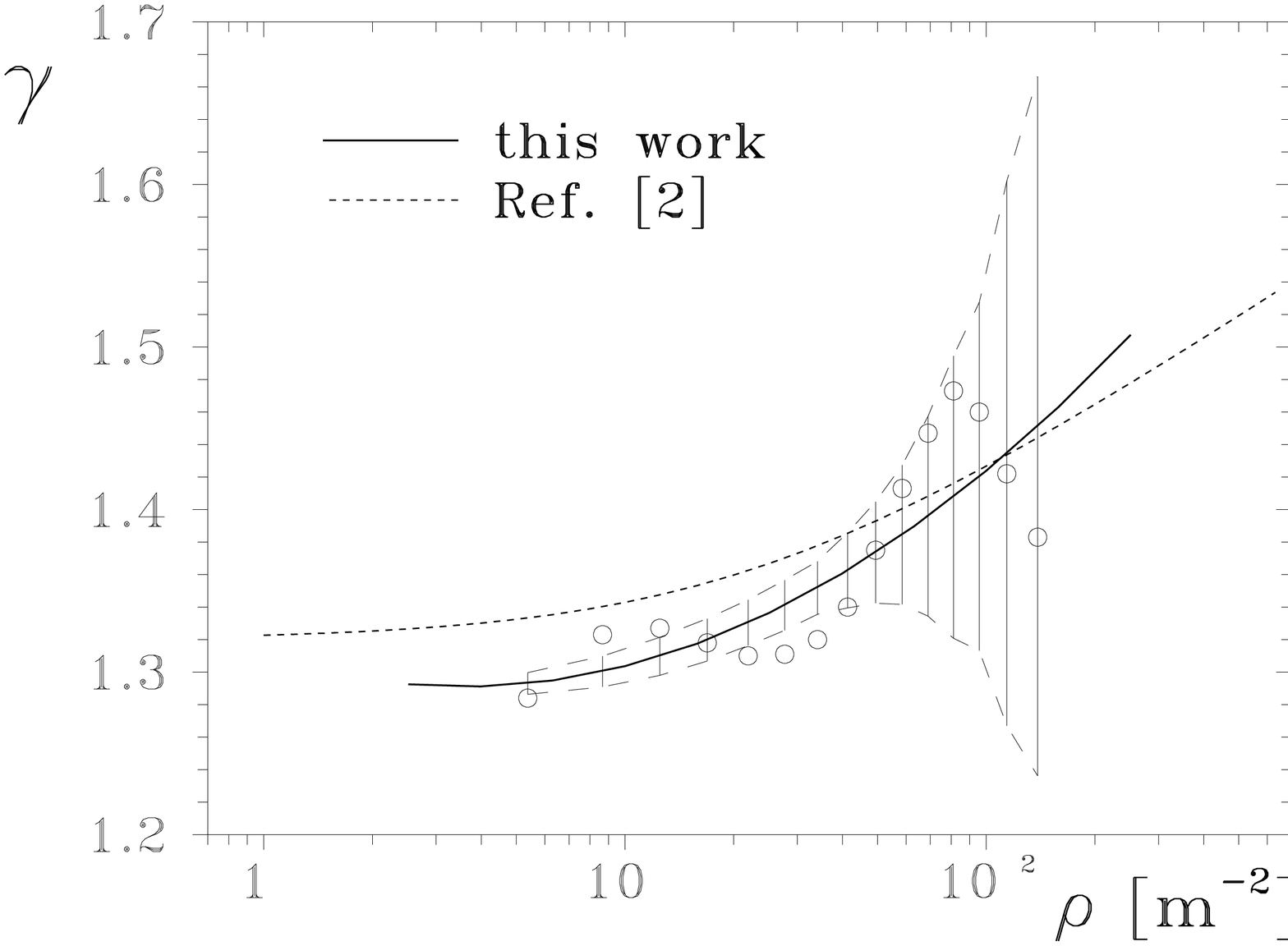,width=10cm}}
\caption{The density spectrum index $\gamma$ as a function of particle
density (the dashed area represents a $1 \sigma$ significance region).
The dashed line is taken directly from the original work of A.
Zawadzki (Ref.\ [2]).}
\end{figure}

Finally the $\gamma(\rho)$ dependence is obtained. It is given in the Fig. 5.
For the comparison the original curve of Ref.\ \cite{az} is also given.
%------------------------------------------------------------------------
%The determination of a confidence level for some functional form adjusted to
%the set of statistically scattered points is not, in general, very obvious.
The dashed area in Fig. 5 is defined as containing the {\em true} dependence
within 68\% probability if its {\em true} form is a second order polynomial
in $\log \rho$. (Small circles presented in the Figure represent
a measured values approximated by the smooth curve (as it is in shown in the
Fig. 2) so the systematics seen is rather a result of the method of the
presentation then a real effect.)
The question of the statistical significance of the change of the slope
is further discussed in Appendix C.
%With such interpretation we can say that the constant value of the spectrum
%index can be accepted only on $\sim 2 \sigma$ confidence level. In other
%words, the probability that the measured results comes by chance from the
%constant index power--law density spectrum is of order of percents.
%------------------------------------------------------------------------

\subsection{The shower size spectrum}
To obtain the shower size spectrum (its index, assuming that the spectrum is in
general of the power--law form) from our data some additional assumptions have
to be made. Shower of size $N_e$ consists of number of particles distributed
over some area thus something like a continuous spectrum of particle densities
appears with respect to the distance to the shower axis for each EAS.
So the Eq.\ (1) has to be modified to:

\begin{equation}
q(\rho) ~=~ \int {\rm d} \overrightarrow{r_0} \int \int
{\rm d} \theta ~{\rm d}\varphi ~
g \big( N_e,~ \theta, ~\varphi \big)~{\left( 1 - {\rm e}^{-sM\rho(r)}
\right)}^2
~~~,
\end{equation}

\noindent
where: $\overrightarrow{r_0}$ is a shower axis position, $\theta$ and $\varphi$ are the
shower zenithal and azimuthal angles (the azimuthal angle dependence is
important here because the G--M counters have a tubular effective
volume and are
separated by some distance one from another in the array of each station).
 The
particle density $\rho(r)$ at the detector site is obviously a function of
the distance from the detector to the shower axis measured in the
shower plane -- $r$ (as it is given explicit in the Eq.\ (5)), but also of
$N_e$ and both inclination angles.
Of course the density
distribution of the particular shower due to the physics of the shower
development is not an unique function of parameters listed above. Its shape
fluctuates about the "mean distribution". Regarding those fluctuations and
making other simplifying assumptions some connections between the density and
size spectra can be found, but the results of such investigations can be
misleading (see e.g. Ref.\ \cite{unknown}).

In the present paper we would like to present results of the full integration
of Eq.\ (5) and then of Eq.\ (3). To proceed with Eq.\ (5) some
knowledge of the EAS
charged particle distribution and its fluctuations is needed. For this purpose
we have used the CORSIKA v 4.12 shower simulation program. The
lateral distribution of electrons from the simulated events has been
parameterized in a form of:

\begin{equation}
r^2~\rho_e(r) ~\sim~ N_e~\left(r \over R_0\right)^\alpha~
\left(1~+~{r \over R_0}\right)
^\beta
\end{equation}

\noindent
for each individual shower. Then the fluctuations from shower to shower
and correlations between the distribution parameters were described
analytically
(to some reasonable extension). Then the Monte--Carlo shower generator was
created which for a given set of ($ N_e,~ \theta, ~\varphi,
~\overrightarrow{r_0})$ values
returns three mean densities of charged particles at each of three G--M
stations positioned exactly as they were forty years ago in \L \'{o}d\'{z}.
The integrals in the Eq.\ (5) were performed by Monte--Carlo and finally the
results presented in the Fig. 6 have been obtained for different constant
values of size spectrum index $\gamma_N$.

\begin{figure}
\centerline{\psfig{file=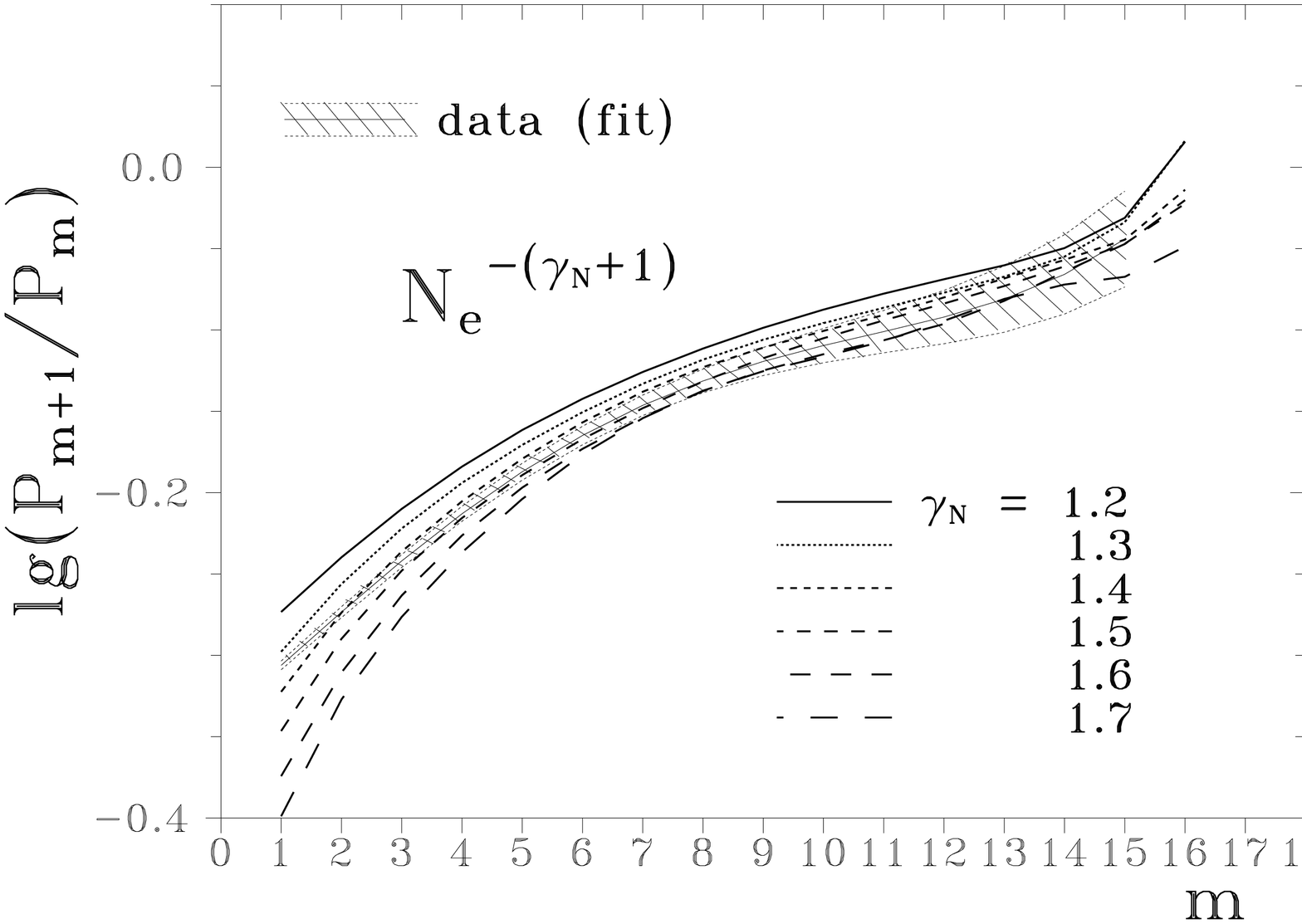,width=10cm}}
\caption{The ratio of $P_{\rm m+1}/P_{\rm m}$ as a function of m calculated
using Eqs.(3) and (5) for different constant values of size spectrum index
$\gamma_N$.}
\end{figure}

The line shows the
experimental result as it was given in the Fig. 2.
%\vspace{1.5cm}

The result is at the first look rather surprising. In comparison with the
density spectrum index the size index variation seen is
statistically negligible ( $1 \sigma$ effect).
%appears to be almost constant.
Further
examinations will be done after the results concerning primary energy spectrum
will be given.

\subsection{The energy spectrum}
The next step is the analysis  of the primary particle energy spectrum.
This needs
another probabilistic function to be assumed. The energy spectrum is to be used
in a convolution with the previously discussed size spectrum and the
probability
density describing the possibility of producing the shower of the size $N_e$
(within d$N_E$) by the primary particle of energy equal to E. Even such a
complex evaluation is greatly simplified in a comparison with the reality. The
primary CR consist of not only one kind of particles. It can be assumed that
there are protons and all the heavier nuclei up to about iron in a comparable
and {\it a priori} unknown proportions (when grouped into five groups: p,
 $\alpha$'s, C--N--O, Ne--Si and $\sim$Fe). The development of a shower
 depends on the nature of the primary particle, so the densities of shower
 particles at the observation level differ in EAS initiated by the CR
particles of the same energy but of different mass.

 The second point we have to face to is that obviously the same shower
 size can be produced by the primary particle of different primary energies
 (even if its mass is the same). One shower can develop deeper in the
 atmosphere (due to probabilistic nature of the process) thus the primary
 particle energy needed to give size of $N_e$ can be smaller than the energy of
 the shower which starts at a very top of the atmosphere. According to this it
 can be expected that the particle density lateral distribution should be no
 longer treated as a function of the shower size (and angles) only ,as it was
 done before in the Eq.\ (5).

 The general formula for a density spectrum evaluation from the primary CR
 energy spectrum can be written as:

\begin{equation}
q(\rho) ~=~ \sum_A \int {\rm d}E
h_A (E)
\int {\rm d} \overrightarrow{r_0} \int \int
{\rm d} \theta ~{\rm d}\varphi ~
g_{AE} \big( N_e,~ \theta, ~\varphi \big)~{\left( 1 - {\rm e}^{-sM\rho(r)}
\right)}^2
~~~,
\end{equation}

\noindent
where: $h_A(E)$ is an energy spectrum of a particular CR particle type and
density $\rho$ is not only a probabilistic function of $N_e$, $\theta$,
$\varphi$
and the distance from the shower core (like it was in the Eq.\ (7)) but also of
the set  ($A$, $E$) values what is a result of the shower development
fluctuations described above.

%Because information about primary mass composition in anyway included in the
%all--particle energy spectrum which is a subject of the present study the
%summation over $A$ in the Eq.\ (7) can be at the moment neglected assuming
%overall energy spectrum $h(E)$. The remaining dependencies on $A$ value (in
%$g(N_e)$, and $\rho(r)$ functions) will be neglected also and the functions
%used further on should be treated as averaged over $A$.
Of course information about primary mass composition is anyway included in the
all--particle energy spectrum.
%which is a subject of the present study the
However the summation over $A$ in the Eq.\ (7) can be at the moment
neglected assuming
overall energy spectrum $h(E)$. The remaining dependencies on $A$ value (in
$g(N_e)$, and $\rho(r)$ functions) will be neglected also and the functions
used further on should be treated as averaged over $A$.

\begin{figure}
\centerline{\psfig{file=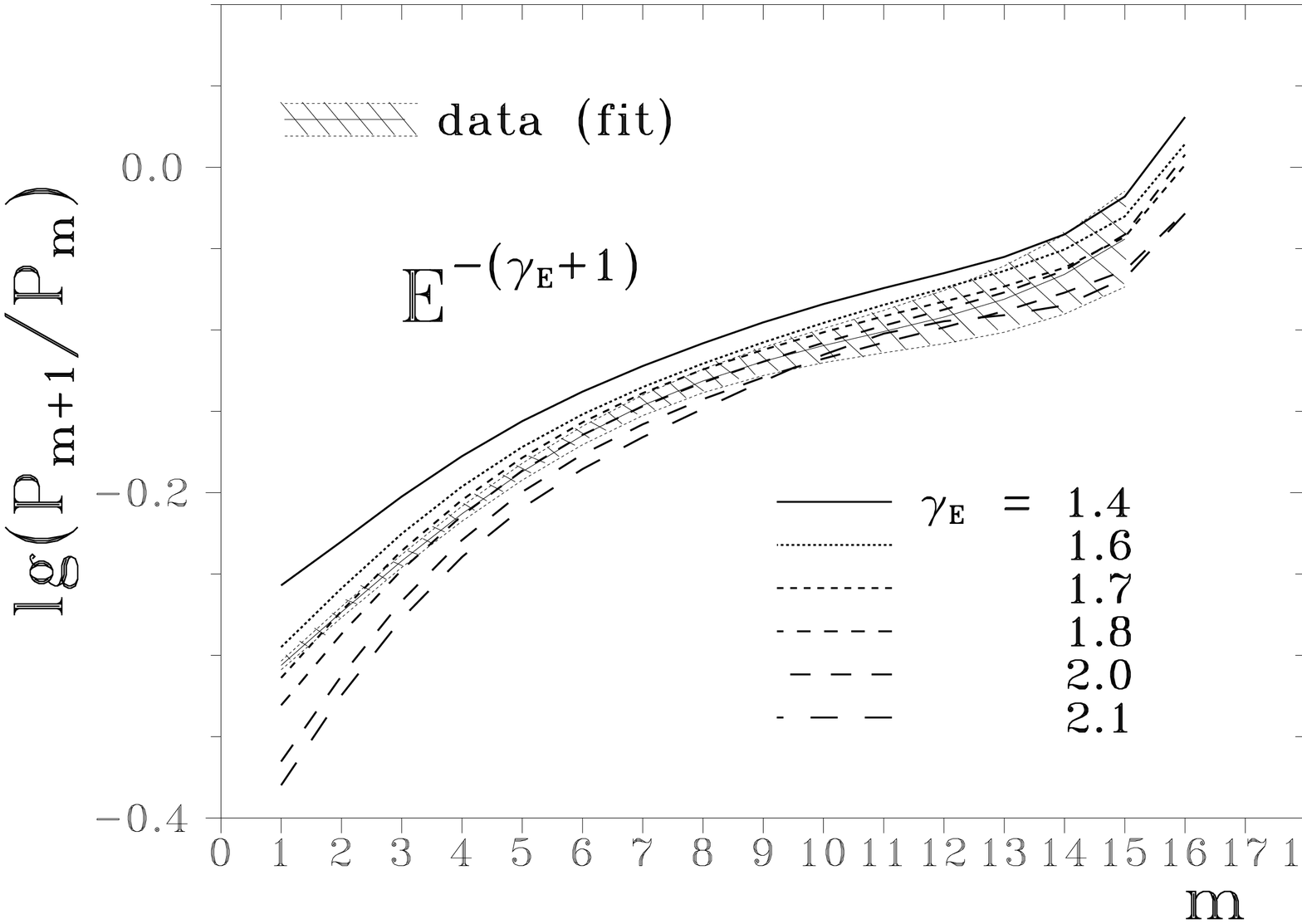,width=10cm}}
\caption{The ratio of $P_{\rm m+1}/P_{\rm m}$ as a function of m calculated
using Eqs.(3) and (7) for different constant values of energy spectrum index
$\gamma_E$.}
\end{figure}

With the help of the CORSIKA simulation program the parameters of the shower
charged particle lateral distribution $\rho$ ($\alpha$, $\beta$ and $R_0$) were
parameterized not only with respect to the $N_e$ but also the rather consistent
description has been found of the value $N_e / \left\langle N_e \right\rangle$
where $\left\langle N_e \right\rangle$ is the mean value of the shower size for
EAS induced by the primary particle of energy $E$. The fluctuations of
$N_e$ with respect to the mean value were also considered.

With all these assumptions the integration of  the Eq.\ (7) was performed by
Monte--Carlo method. Then using the Eq.\ (3) the results presented in
the Fig. 7
were obtained for different values of the index $\gamma_E$ of the primary CR
energy spectrum $h(E)~\sim~E^{-(\gamma_E+1)}$.

The line shows the
experimental result as it was given in the Fig. 2.
%\vspace{1.5cm}

The results are very similar to the one obtained for the size spectrum
evaluation. The change of the power--law index, substantial for the density
spectrum, is not seen for the energy spectrum as well.

\section{Discussion}

\subsection{Size and energy spectra}

First we want to show quantitatively the reason of the disappearing of the
change of the power--law indexes $\gamma_N$ and
$\gamma_E$ while the index change for the density spectrum, $\gamma$,
is unquestionable.

The Figures 8 (a) and (b) are analogous to the Fig. 4. They present the
distributions of shower sizes (a) and primary energies (b) contributing to the
registration of a particular number of m G--M tubes registered charged
particles in the event.

\begin{figure}
\centerline{\psfig{file=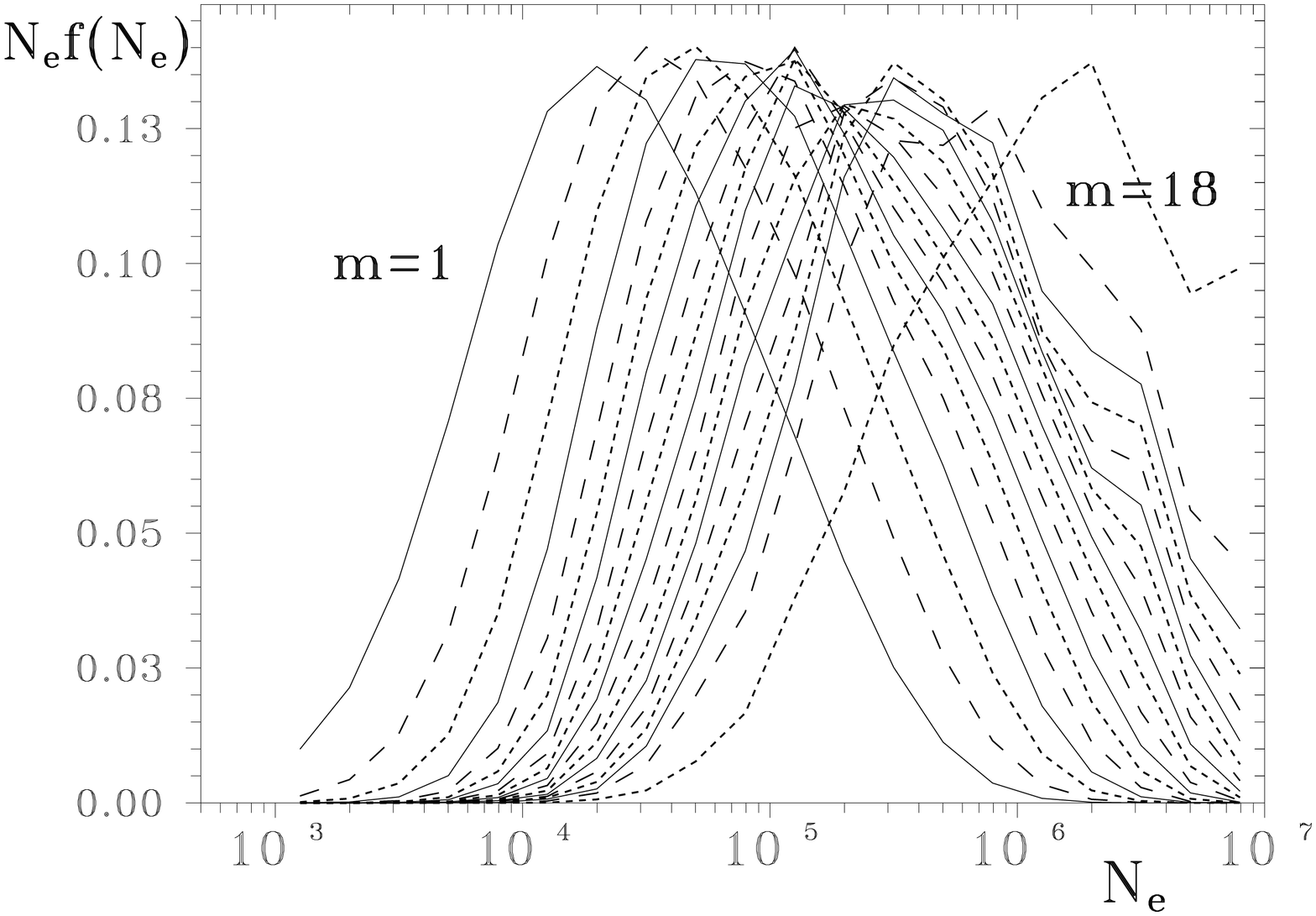,width=7cm}
\psfig{file=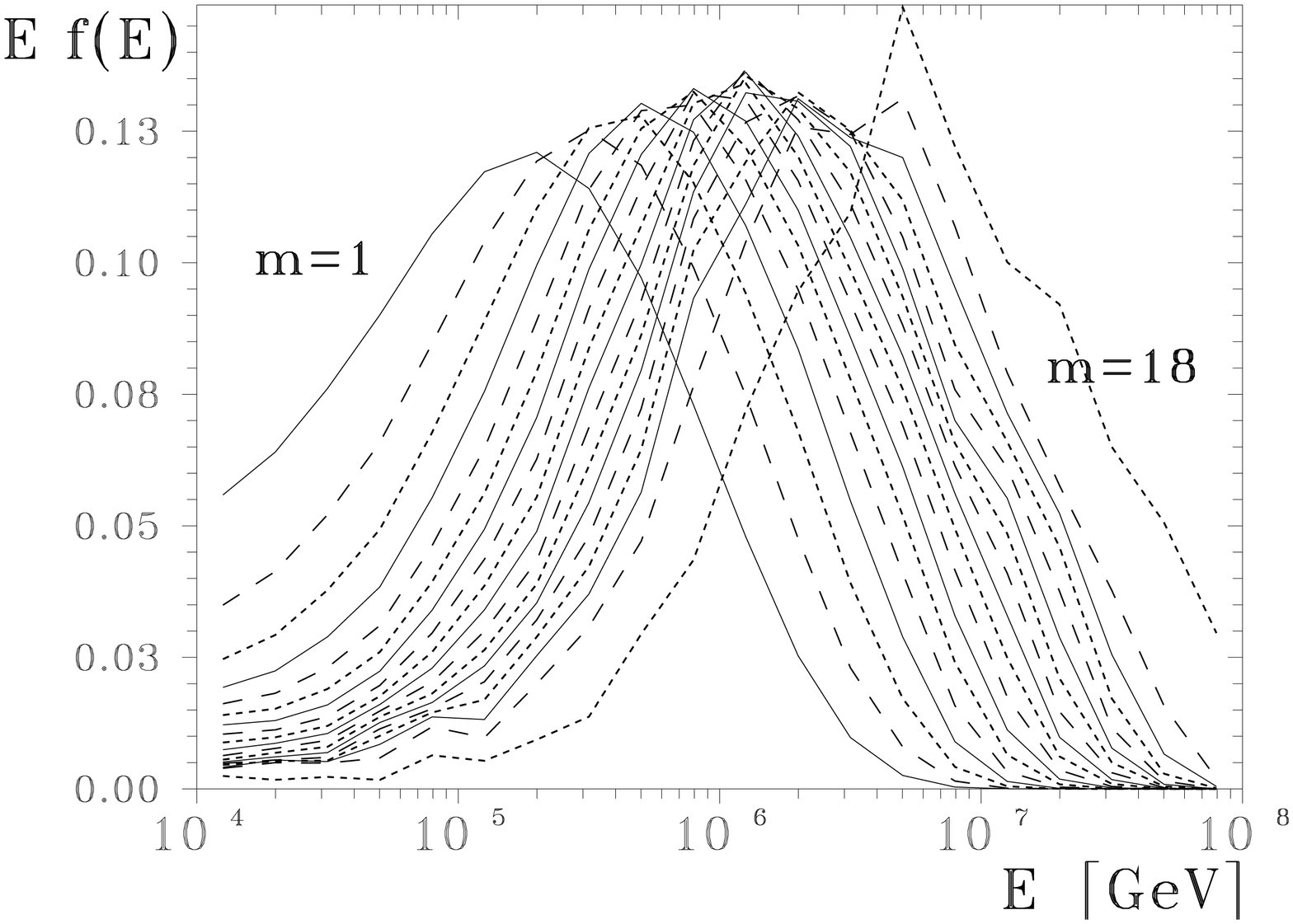,width=7cm}}
\caption{The distribution of the shower sizes $N_e$ (a) and primary energies
$E$ (b) giving a contribution to the m-hit rate.}
\end{figure}

In contrast to the Fig. 4 where the distributions were rather well separated,
the
distributions in the Fig. 8. are very much wider and overlapping. In the Fig.4
the width
(FWHM) of the density distribution in the case of m = 7 is comparable with the
distance between the mean values of $\rho$ for m = 7 and m = 10 while in the
Fig. 8a the FWHM for m = 7 is of order of the distance between m = 2 and m =
16 . For the energies (Fig. 8b) the situation is even worse.
This according to the integration procedure leads to smearing out all
details in the generally very fast falling spectra.

Due to the large widths of the distributions in the Fig. 8 it is not reasonable
to calculate the mean values of $N_e$ and $E$ responsible for the spectrum
slope at given m--hit rate as it has been done in the Fig. 5 for the density
spectrum. However it is interesting to note that the whole range of sizes and
energies for m $\ge$ 3 is of about one decade. It is clear that any change in
the spectrum slopes which could be seen by the experiment has to be
rather sharp and substantial.

\subsection{Muon density spectrum}

The interesting possibility is to use the density spectrum method discussed in
this paper for a muon component of EAS. Muons are known to be stronger
correlated to the primary particle energy and number of muons in the shower
fluctuates weaker than the electron shower size (what is due to the difference
in their origin mechanism and the shower development geometry). With the help
of large EAS array which can be subdivided into smaller parts each giving
separately a kind of hodoscopic information about the muon component in the
shower the statistic needed can be accumulated relatively fast.
%, much faster than one needed for the conventional method of
%analyzing the muon content of each shower or shower group.

The experiment KASCADE (Ref.\ \cite{kascade}) gives an excellent opportunity
for such
studies. Its muon array part consists of 192 detector stations each of them
equipped with four plastic scintillators of area of ${\rm 0.81}$ ${\rm m^2}$
placed
below the lead and iron shield which stands as a soft shower component
absorber. The minimum muon energy required is equal to 300 MeV. Scintillators
in every station are viewed by four photomultipliers in such a way that each
PM sees two scintillators and the data acquisition system collects
hodoscopic information about status of every PM tube. Thus there is a huge
amount of ``binary'' information about every event not disturbed by
fluctuations common to ``analogous'', pulse--height measurements.
This information can be used to obtain a very precise and accurate
results concerning the muon density spectrum. The data analysis is in progress
and results will be published soon.

The studies of muons in
EAS are very important from the point of view of primary mass composition
determination. Difficulties concerning estimation of total muon content in
individual showers (due to low densities of particles) make the muon density
spectrum measurements rather attractive.

\section{Conclusions and Summary}
The forty years old data were reexamined using more powerful computational
methods available at present. The confirmation of the original work
(Ref.\ \cite{az}) results is given. Further evaluation of the same data set
shows
the general agreement with the power--law spectra of EAS sizes and primary
particle energies.

For the extremely small showers there is seen a substantial change of the
assumed size and energy spectra shape. However, there are (at least) two
undistinguishable possibilities of explanations of that experimental fact.
%The
%first is that the particle contents of the very small showers decreases
%drastically faster than predicted by the Monte--Carlo simulation program with
%decreasing of the primary particle energy.
The one is the truncation of the fluctuation distribution of the
particle number seen by the limited area detector.
This needs to be studied carefully from the point of view of the
earth surface EAS experiments intended to fill the gap between the
balloon--borne direct measurements and existing ground level
experimental results concerning the
nature of the primary cosmic ray spectrum. The other interesting aspect of the
problem is the exactness of the comparison of small EAS measurements by the
arrays of different geometries (and thus triggers) and positioned at different
altitudes.

The standard investigation of such small showers is difficult due to very
small particle number, so large areas of relatively sensitive detectors with
negligible background noise are needed. The method of the density spectrum
measurement by the small hodoscopic counters could be a solution here. To go to
smaller energies the new measurement has to be performed with the geometry
chosen carefully to get a maximum efficiency in the region of interest.

From the other side, for the density spectrum method the increase of the
primary
particle energy under study will verify or, at least, confirm in less
''Monte--Carlo'' dependent way the results about the ''knee'' problem still
discussed by many authors both from experimental and interpretational point of
view.
The density spectrum method for a muon component of EAS can be used for such
purposes.

Concluding we would like to summarize the present work in a few sentences:

\begin{itemize}
\item{ Charged particle density spectrum in the range of 1 $\div$ 100
${\rm m}^{-2}$ is power--law with the index changing slowly from 1.3 to 1.5.}
\item{ Size and primary particle energy spectra were found to be power--law
with the indexes $\sim$ 1.5 and 1.8 respectively and no significant
change of these indexes is seen (see Appendix A).}
\item{ Further examination of the small showers behaviour is needed to explain
some features seen by the experiment. The density spectrum method can be
useful for such investigations (see Appendix B).}
\item{ Measurement of the muon density spectrum can give a new interesting look
into the problem of the energy spectrum near the ''knee'' region
($\sim 10^{15}$ eV).}
\end{itemize}

\appendix

\section{Density spectrum measurements; historical remark}

The situation prior to 1957 was summarized by Greisen in Ref.\ \cite{gra}.
All measurements show the slight and constant increase of the index in
power--law
form of the density spectrum. This, in fact, implies that the density spectrum
has not a straight power--law form. The measurement of Zawadzki was the first
accurate enough which allows one to perform more detailed analysis of the
spectral index behaviour (Ref.\ \cite{coc}). The point was that obtained
change of the index was rather sharp. For densities smaller than few tens of
particles/${\rm m^{2}}$ it seems to be constant and the possibility of
significant change was indicated for higher densities. This experimental
fact gave an assumption
to treat the CR spectrum as having a broken power--law form with one index
below a certain point and much steeper above. Implications of such a picture
are rather far--going and no matter how the original Zawadzki measurement
confirmed these speculations the idea of a ``knee'' or two--component CR
spectrum arise and it is still one of the most important features of the
cosmic ray phenomenon nowadays.

The paper with the Zawadzki experiment results was originally published
in French and distributed not very widely. Confirmation of this fact
can be found e.g. in Ref.\ \cite{allan} where still in 1979 year the old
Greisen (Ref.\ \cite{gra}) summary of the world data on density spectrum
are given without the Zawadzki result. The problem is, in fact, not in omitting
the most significant measurement but in referring to the points which are known
to be not exactly correct due to some simplifications made during the data
analysis. In Ref.\ \cite{az} it was shown that results obtained
in Ref.\ \cite{coco} should be corrected. Anyhow rather limited
statistics
of this experiment situate the proposed corrections inside error bars of the
original points. For the same reason the other measurements (Refs.\
\cite{broa,hods}) when reexamined, do not contradict the Zawadzki paper conclusions
about a sharp change of density spectrum index.

%\appendix

\section{Low density part of the spectrum}

We want to discuss briefly in this appendix
the clear discrepancy with the constant slope of
energy and size spectra seen for first 2 or 3 points in the Figs. 6 and 7.

First it has to be mentioned that this can not be explained by the obvious
flattering of the particle lateral distribution with decrease of the shower
size
(thus the primary energy). Smaller showers become ``older'', so the continuous
change of their lateral distribution is expected and was confirmed in the
simulations (using CORSIKA code). This can be a reason of the
difference between the size (energy) and density spectra seen in
Figs. 3 and 6. However, the effect of the continuous change of the particle
lateral
distributions in EAS was taken into account in the respective integrals
calculations for $N_e$ and $E$ spectra -- and the sharp decrease of
$\gamma_N$ and $\gamma_E$ for m = 1,2 still exists there.

The one most obvious possibility is that the charged particle lateral
distribution obtained via CORSIKA program does not correspond to the reality
for
a very small shower. This way of explanation can explain of course everything,
but unless we have no other reason for claims about the CORSIKA incorrectness
we have to be careful in such suppositions.

Another possible explanation is connected with the particular shape of the
particle number fluctuations at the certain (small) area in the shower plane
for very small showers.

This is closely related to the other relatively old experimental effect
reported in Ref.\ \cite{dkg}. The Authors presented there data
taken at Cornell University by recording the pulse height distribution
in a single scintillator of area ${\rm 0.86 m^2}$ with and without the
requirement that a single particle be detected simultaneously in
another counter a few meters away. The data clearly show that while for the
high particle densities the density spectrum has a well defined power--law
behaviour with the index of --1.62 then for smaller densities the slope
seems to be also -- 1.62 only if the additional coincidence is required.
With the single detector trigger the slope changes dramatically up to --3.19.

%It is very interesting that after about
%30 years the same effect was reported at the second Moscow
%International Cosmic Ray Conference
%by some people of the ''Baksan carpet'' group \cite{baksan}.
%Authors claim that
%''it appeared that a longstanding {\em mystery} had been solved''
%by introducing the {\em new physics} called {\em narrow showers}. However
%the Rapporteur of the Adelaide
%International Cosmic Ray Conference
%(Ref.\ \cite{reid}) seemed not to be
%fully convinced by their explanation saying ''It seems that experimentalists
%cannot yet be sure that they fully understand such a commonplace feature of
%detector response.''

The careful statistical analysis performed during our density spectrum studies
allows us to solve ``that longstanding mystery'' \cite{reid}.
In the calculations presented
above it was assumed that the number of particles
fluctuates in a common poissonian way about the mean value given by the
particle density times the detector area factor. Assuming that the density
spectrum is exactly the power--law one (with, e.g., $\gamma$ = 1.3) we can
consider the particle number spectrum seen by the unique area detector.
Its distribution is given by

\begin{equation}
P_{\rm n}~=~ \int_{0}^{\infty} {\rm d}\rho~\rho^{\gamma +1}
~p_{\rm n}(\rho)~~~.
\end{equation}

\noindent
where $p_{\rm n}(\rho)$ describes the spread of number of particles seen
with respect to the mean value $\rho$.

The results for scarcely different $p_{\rm n}(\rho)$ distributions
are given in the Fig. 9 (divided
by the incoming density spectrum to better see the disturbance produced
by the fluctuations).

\begin{figure}
\centerline{\psfig{file=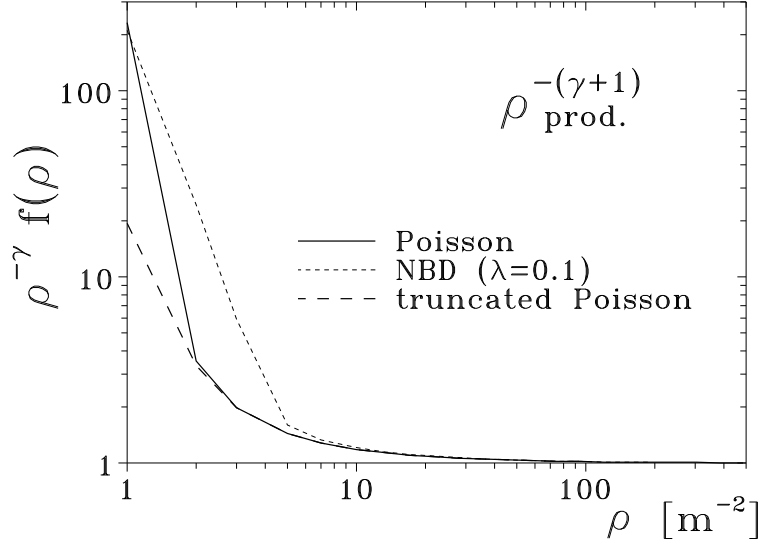,width=10cm}}
\caption{The spectrum of the number of particles seen in the detector area
divided by the density spectrum of the form $\rho ^{-(\gamma +1)}$
(with $\gamma$ = 1.3). The solid line is for pure poissonian fluctuations,
the short dashed line is for the Negative Binomial Distribution with the
parameter $\kappa$ = 0.1 and the long dashed line represents the truncated
Poisson fluctuation.}
\end{figure}

Curves in the Fig. 9 correspond to the Poissonian distributed
fluctuations (solid line)

\begin{equation}
p_{\rm n}(\rho)~=~ {{\rho ^ {\rm n}} \over {{\rm n}\!}}~{\rm e}^{-\rho}~~,
~~~
\end{equation}

\noindent
the slightly wider distribution of particle number fluctuations
as given by the Negative Binomial Distribution (NBD)
(short dashed line)

\begin{equation}
p_{\rm n}(\rho)~=~ {{\rho \kappa+{\rm n}-1} \choose {\rm n}}
\left[ {{\kappa }\over
 {1+\kappa}}
\right]^{\rho \kappa}~\left[ {1 \over  {1+\kappa}}
\right]^{\rm n}~
~,
\end{equation}

\noindent
and the Poissonian distribution truncated at 20$\sigma$ (possibility of
registration of m particles when the expected number is $\overline \rho$
is 0 when m$> \rho \: + \:20 \sqrt{\overline \rho}$).
The differences are rather small and occur mainly
in very tail regions. However, due to the steep
density spectrum the tails are extremely important, as it is seen in the
Fig. 9. Even the poissonian tails lead to the artificial overabundance of the
small particle numbers.
%Thus we see that the explanation of the Cornell University experiment
%does not need any additional hypothesis like e.g. ``narrow showers'' proposed
%in \cite{baksan}.

The differences are not so substantial
in the Zawadzki experiment like it is presented in the Fig. 9
due to the existence of the three-fold coincidence.
This effect was exactly taken into account in our calculations (Eq.\ (4)).
However, it is important to note that even a small change
of the fluctuation shape can
lead to the change of the expectations for very small showers.

The widening of the fluctuation distribution
leads to the more pronounced overestimation of the small particle number rate.
%This effect enlarged by additional fluctuations of the scintillation
%counter signal which are of the Landau type is exactly what was seen in
%the Cornell University and Baksan carpet measurements.
The vanishing of the
effect with additional coincidence requirement is thus obvious.

Coming back to our main subject, the Zawadzki measurement interpretation,
one can see that the rate of events containing
small particle numbers looks to be smaller.
On the first sight it is hard to imagine that the particle number seen in the
detector can fluctuate narrower than poissonian. However, one has to remember
that we have to take into account a very steep density spectrum so the single
or two--particle events could occur as a result of large fluctuations
for relatively small mean values.
%(which appears in the density spectrum much more frequently).
The character of the Poisson distribution is
such that there is still a possibility of recording an event with any high
number of particles for any small expected value. The physical constraints,
however, convince one that the unusually high particle numbers simply must not
be produced by small showers. Thus, the first approximation poissonian
distribution should be truncated at some (distant) point, anyhow.
This of course
does not disturb the mean values and the spectra for high densities, but can
effect the integration in Eqs.\ (8) and (4) for small values of m in the way
that the calculated above ratio ($P_{\rm m}$) overestimate the m--hit ratios
for small values of m as it is shown by the long dashed line in Fig. 9.

\section{Statistical significance of the change of slope of the
density spectrum}

The analysis of the change of the density spectrum slope presented above
has been made using ratios of registered abundances of m and (m+1) G-M
counters signals. This method assured the continuity of the density
(differential) spectrum while changing the slope parameter. In other words
we have used the derivative of the spectrum while comparing the
experiment with calculations.
However when trying to answer the question about the statistical
confidence of the measurement the whole experimental information ought
to be used (like it is presented in Fig. 1).

The most obvious way of the determination of the correctness
of the theoretical description of a measurement is to use the $\chi^2$
test. With the data given as a simple numbers of counts it is straightforward.
Doing the textbook calculations the values of $\chi^2 / NDF$
equal to 1044, 417, 48, 80, 573, 5130, and 18500
have been obtained for the density spectrum indexes $\gamma$
of 1.2, 1.25, 1.3, 1.35, 1.4, 1.5, and 1.6, respectively.
Of course all of them are extremely unacceptable on any confidence level.
To understand better the origin of such exceptionally high values
a short clarification is needed.
It would be at least a misinterpretation to
use the numbers given above to determine the statistical significance
of the statement on the change of the slope of the density spectrum,
specially in connection with the discussion in the preceding Appendix.
(One can ask also about the puzzling inconsistency between these values and
results given, e.g., in Fig. 3.)

Some useful information can be found investigating contributions to the
overall $\chi^2$ coming from different values of m. If we define the
function $\chi^2(m)$ as

\begin{eqnarray}
\chi^2(m)/NDF~=~
{{\sum_{i=m}^{17} (N_i~-~N\: P_i)^2 / N_i}
\over {17~-~m}}~,~\nonumber \\
~~{\rm where}~~~~N~=~{{\sum_{i=m}^{17} N_i}\over{\sum_{i=m}^{17} P_i}}~~~;
~~~~~(m~<~17)~~,
\end{eqnarray}

\noindent
then respective contributions can be compared. Results of calculations
are presented in Fig. 10.

\begin{figure}
\centerline{\psfig{file=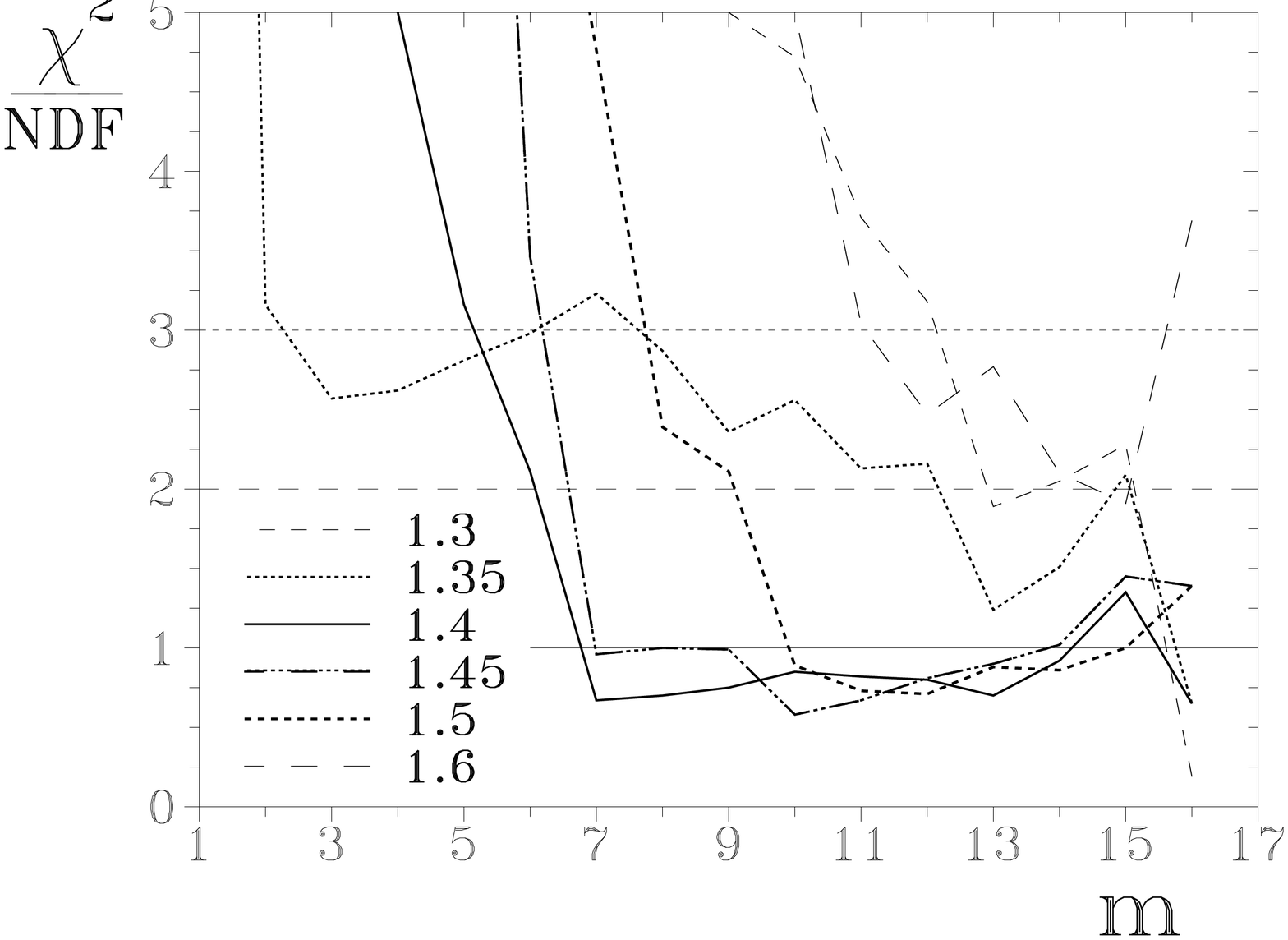,width=10cm}}
\caption{Contributions to the overall $\chi^2$ coming from different
parts of the density specrum (m). Lines connect points calculated for
integer m for different values of the desity spectrum indexies
$\gamma$.}
\end{figure}

Few important features can be seen.
First is that only for the values
of index $\gamma$ in the range $\sim 1.4~ \div~1.5$ $\chi^2/NDF$
of order of 1 can be achieved in some (limited to m $>$ 7 $\div$ 9)
density range. The second point is the extraordinary rise of
$\chi^2$ for very small values of m. This is the statistical
confirmation of the problem of small density discussed in the
Appendix B.
The Fig. 10 shows clearly the necessity of the change of the
spectrum index from about 1.45 above m=7 to a smaller value (which,
however, can not be concluded form such $\chi^2$ analysis).
Thus the inconsistency of the overall $\chi^2$ with Fig. 3 seems to be
apparent.

The $\chi^2$ analysis can be of course performed for the shower size
and primary energy indexes ($\gamma_N$ and $\gamma_E$). Respective
results are given in Fig. 11 (a) and (b).

\begin{figure}
\centerline{
\psfig{file=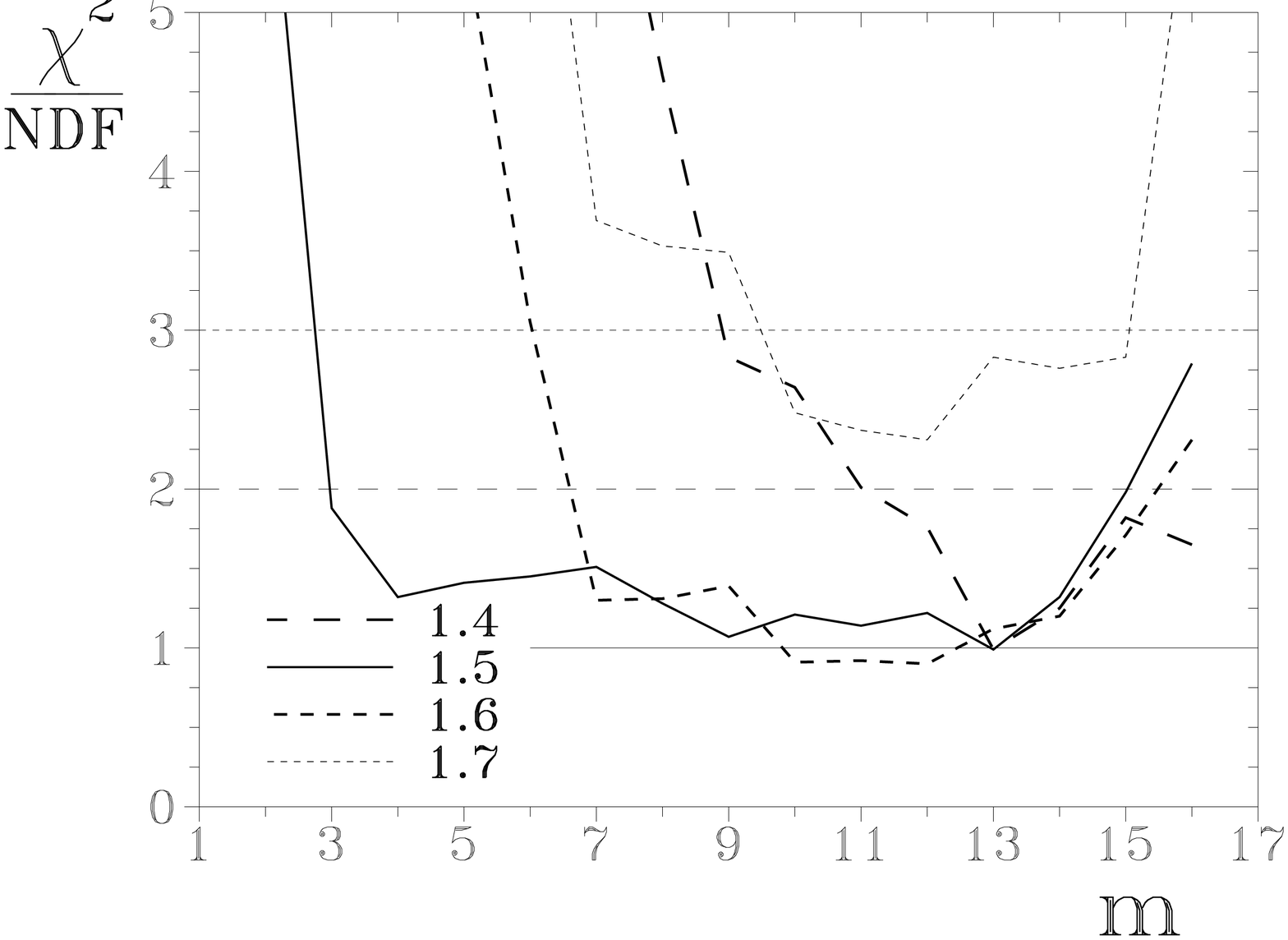,width=7cm}
\psfig{file=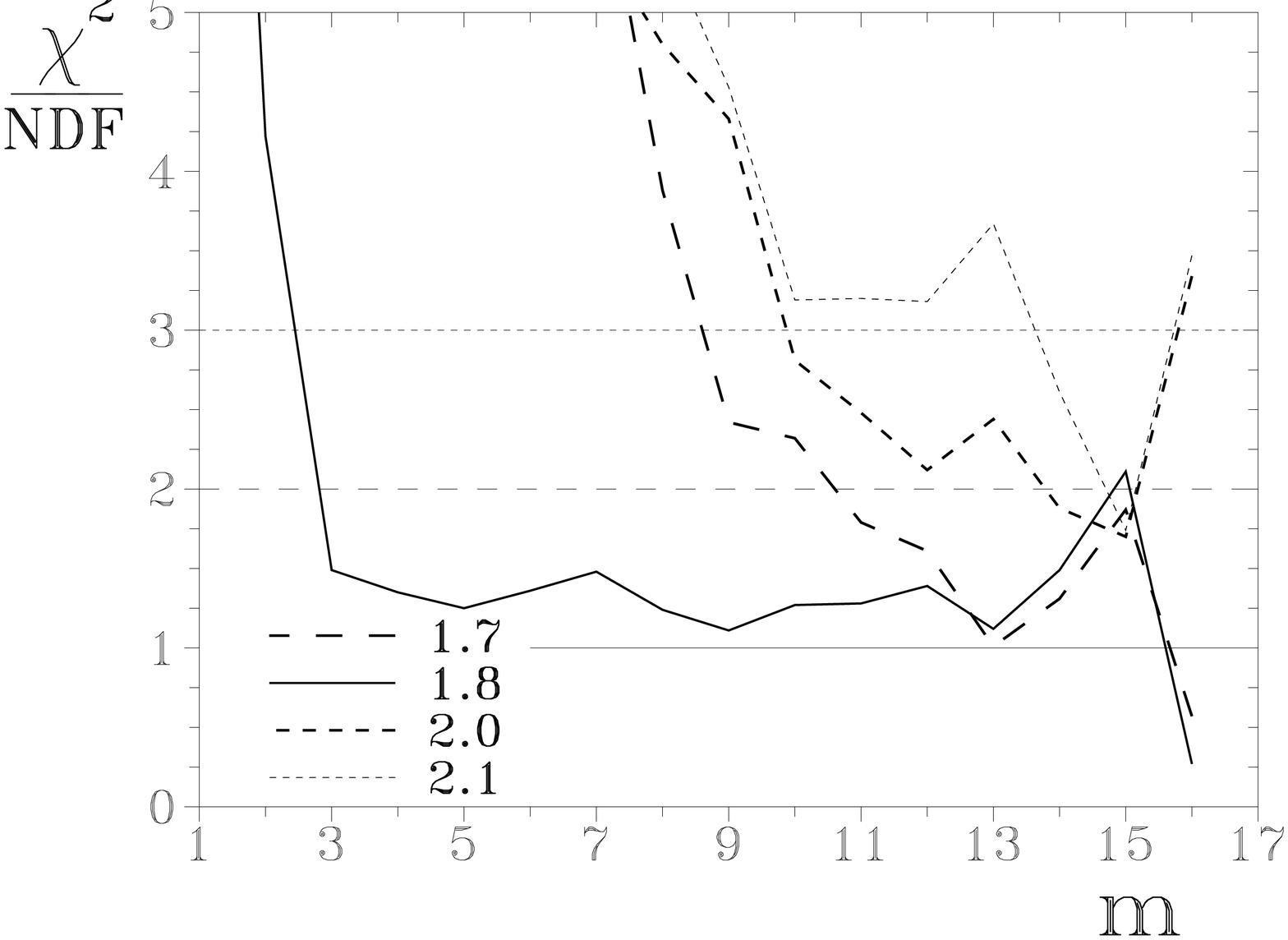,width=7cm}}
\caption{Contributions to the overall $\chi^2$ coming from different
m values for shower size (a) and primary energy (b). Lines connect
points calculated for
integer m values for different desity spectrum indexies
$\gamma_N$ and $\gamma_E$.}
\end{figure}

Again the overall values of $\chi^2$ are extremely large, but the
plotted different m contributions confirmed our conclusion
made on a basis on Figs. 6 and 7 that the both size and energy spectra
are not in contradiction (except of m = 1, 2) with the single power--low
spectral index of $\sim$ 1.5 and $\sim$ 1.8, respectively.

\end{document}